\documentclass[aps,pra,twocolumn,floatfix,superscriptaddress]{revtex4}  
\usepackage{float}
\usepackage{graphicx}
\usepackage{amssymb}
\usepackage{dsfont}
\usepackage{amsmath}  
\usepackage{epstopdf}
\usepackage{color}

\def\ket#1{|#1\rangle}

\def\w{\omega}

\def\wp{\omega^\prime}

\begin{document}

\title{\bf Analytical results for a conditional phase shift between single-photon pulses in a nonlocal nonlinear medium}
\author{ Balakrishnan Viswanathan and Julio Gea-Banacloche}
\date{\today}



\begin{abstract}
It has been suggested that second-order nonlinearities could be used for quantum logic at the single-photon level. Specifically, successive two-photon processes in principle could accomplish the phase shift (conditioned on the presence of two photons in the low frequency modes) $ |011 \rangle \longrightarrow |100 \rangle \longrightarrow -|011 \rangle $. We have analyzed a recent scheme proposed by Xia et al.  to induce such a conditional phase shift between two single-photon pulses propagating at different speeds through a nonlinear medium with a nonlocal response. We present here an analytical solution for the most general case, i.e. for an arbitrary response function, initial state, and pulse velocity, which supports their numerical observation that a $\pi$ phase shift with unit fidelity is possible, in principle, in an appropriate limit.  We also discuss why this is possible in this system, despite the theoretical objections to the possibility of conditional phase shifts on single photons that were raised some time ago by Shapiro and by one of us.
\end{abstract}
\maketitle

\section{Introduction}

It would be extremely useful for all kinds of quantum information processing if one could use optical nonlinearities to implement quantum logical gates at the single-photon level.  The idea of using the cross-phase modulation properties of an optical Kerr medium (a $\chi^{(3)}$ nonlinearity) to change the overall sign of the wavefunction when two photons (as opposed to one) are present together in the medium (thereby carrying out a basic entangling operation known as the ``conditional phase'' or CPHASE gate), was first popularized, very early on, by Chuang and Yamamoto \cite{chuang}, and actually predates, in a somewhat different form due to  Milburn \cite{milburn}, much of the current interest in quantum computing.  There are, however, considerable practical difficulties to its realization, due primarily to the fact that conventional optical nonlinearities are extremely weak at the single-photon level.  

Interestingly, there have also been a number of theoretical papers, primarily by J. H. Shapiro and coworkers \cite{shapiro1,shapiro2,dove}, as well as one of us \cite{jgb2010,GeaBNeme14,ViswGeaB15}, and others \cite{HeSimon,BingHe,xu}, strongly suggesting that there are some fundamental obstacles to the direct realization of high-fidelity conditional phase gates using any kind of optical nonlinearity.  Typically, what is found in papers such as \cite{shapiro1} and \cite{jgb2010} is that there is a tradeoff: large phase shifts are associated with low gate fidelities, and vice-versa.

Recently, however, several developments have come to challenge the skepticism expressed in the above-mentioned works.  
In particular, at least a couple of theoretical papers \cite{knight,brod1} have appeared that make a strong case for the possibility of a unit-fidelity $\pi$-radian conditional phase shift in different nonlinear optical setups, in more or less direct conflict with the expectations of \cite{shapiro1,shapiro2,dove,jgb2010,GeaBNeme14,ViswGeaB15,HeSimon,BingHe,xu}.  (Other encouraging theoretical results along these lines have been presented in \cite{chudzicki,niu}.)  At the same time, on the experimental side, several groups have reported impressive conditional phase shifts at the single-photon level over the past year, using different arrangements, such as storage and retrieval of the photons in an atomic medium \cite{beck,tiarks}.  All this suggests that a re-examination of the real limitations of traveling photon schemes may be particularly timely (if not, in fact, somewhat overdue). 
    
As a first step towards this goal, we carry out in this paper a thorough analytical study of the system studied numerically by Xia et al.~in \cite{knight}, namely, two single-photon pulses co-propagating, at different velocities, through a nonlinear ($\chi^{(2)}$-equivalent) and nonlocal medium, in its most general form.  Our results confirm their numerical observation that, indeed, a $\pi$ phase shift with unit fidelity is possible, in principle, in this system, in an appropriate limit that we characterize here.  Equally importantly, we also discuss how this is possible, in spite of the theoretical difficulties raised in the works mentioned above.  

Our paper is organized as follows. In Section II, we summarize the main difficulties pointed out in the original papers by Shapiro \cite{shapiro1} and Gea-Banacloche \cite{jgb2010}, and pointing out how, in fact, one could get around these problems, and how this is accomplished in the proposals of Xia et al. \cite{knight} and Brod and Combes \cite{brod1} (see also \cite{brod2} for a more detailed discussion of the latter).  Then, in section III, we introduce our generalization of the system of Xia et al., present the formal solution to it, and discuss analytically the large-phase shift, high-fidelity limit.  Section IV is devoted to a more qualitative, conceptual discussion of what makes this scheme work.  Section V summarizes our conclusions and the questions that we feel still remain to be addressed.  Finally, an Appendix shows how to obtain an approximate analytical solution for the time-evolution of the single-photon wavefunctions in this system, under the conditions leading to the maximum fidelity and phase shift.

\section{Obstacles to high-fidelity single-photon nonlinear optics}
The main obstacles to achieving high-fidelity in nonlinear optical processes at the single-photon level identified in refs. \cite{shapiro1} and \cite{jgb2010} are:
\begin{itemize}
\item{($i$)} Phase noise arising from the Langevin operators introduced to preserve unitarity in a $\chi^{(3)}$ process, when the medium response is nonlocal in time \cite{shapiro1}.
\item{($ii$)} Spectral entanglement of the output photons, due to the fact that the original photons are destroyed and recreated with the sole constraint of overall energy conservation  \cite{jgb2010}.
\end{itemize}
Mathematically, the origin of the first of these difficulties is ultimately the need to limit the medium's bandwidth in order to avoid infinities.  Physically, the finite bandwidth is naturally tied to a finite response time for the medium, which one can always expect to be the case for any realistic system.  As pointed out already in \cite{haus} (see also  \cite{blow,kartner,joneckis}) ``quantum theories of light in instantaneous Kerr media are ill-defined in the absence of dispersion due to the infinite bandwidth of the vacuum fluctuations coupling to any frequency window of interest. Hence, even though one is interested in the quantum evolution of long pulses, reference to the much shorter response time of the non-linearity is unavoidable.'' 

The introduction of this response time in the evolution equations for the quantized field, when done in the standard way (namely, by giving a ``memory'' to the nonlinear index of refraction, as in \cite{shapiro1}) has, however, a profound consequence.  The resulting time-nonlocality means that the system's evolution cannot be derived from a Hamiltonian that contains only free-field operators.  This, in turn, means that the free-field commutators will not be preserved (and hence, unitarity will be violated), unless appropriate ``Langevin noise terms'' are added.

The effect of these noise terms is negligible in the case when the pulses are very long compared to the medium's response time, since in that case the latter can, for practical purposes, be treated as instantaneous.  This is the ``large medium bandwidth'' regime.  However, in this limit Shapiro found that the optical nonlinearity effectively vanishes, so the cross-Kerr phase shift goes to zero.  This may be understood as arising from the fact that, if we only have two photons in a very long pulse, the probability that they could both (randomly) be found within the same narrow time window corresponding to the medium response time becomes negligible. 

In a similar way, one can expect that the pairing of a narrowband medium with a broadband wavepacket (i.e., the very short-pulse limit) will be useless, because the medium will either reject (that is, reflect) or absorb most of the incident spectrum; in the first case, with high probability the photon state is unchanged, and in the second case the photons simply vanish.  So the only potentially useful regime is when the medium bandwidth and the pulse bandwidth (or equivalently, the pulse duration and the medium response time) are more or less evenly matched.  In this case, however, the effect of the Langevin noise operators cannot be neglected, and one finds that the fidelity of the conditional gate is substantially degraded: that is to say, the outgoing pulses do not overlap very much (either spectrally or temporally) with the incoming ones. 

The considerations above originated in the study of conventional nonlinear devices, such as crystals that are intended to be used far from resonance, and can therefore be characterized by a real (nonlinear) index of refraction.  However, in the years preceding the publication of Shapiro's paper, many workers in quantum information had considered near-resonance pulse propagation in especially configured atomic media, which led to evolution equations (typically written in the Heisenberg picture) that mimicked, at the single-photon level, what one would obtain for classical fields propagating through a Kerr medium.  In an apparent contradiction with the claims of \cite{shapiro1}, these equations were unitary and local in time, and did not include explicitly any Langevin noise terms.  

In an attempt to understand which limitations, if any, might be present in these systems, one of us \cite{jgb2010} developed a Hamiltonian treatment of the so-called ``giant Kerr effect'' \cite{ima,lukin,rmp} and presented its solution in the Schr\"odinger picture.  The main obstacle to high-fidelity performance that was revealed by this study was item ($ii$) above, that is, the spectral entanglement of the output photons.  Its origin is simple: in these systems, to get a large phase shift the photons need to interact very strongly, which means that the input photons are actually (and not just virtually) destroyed and then re-created in the medium.  However, in the co-propagating case with equal velocities, the only constraint that applies to the whole process is conservation of energy, which is equivalent, in this configuration, to conservation of momentum or phase matching, and which only restricts the outgoing frequencies $\omega_1,\omega_2$ to satisfy the relation
\begin{equation}
\omega_1 + \omega_2 = \wp_1 +  \wp_2
\label{ne1}
\end{equation}
where $\wp_1,\wp_2$ are the frequencies of the incoming photons.  Equation (\ref{ne1}) typically leads to an entangled spectrum of the form 
\begin{equation}
f(\omega_1,\omega_2) \sim \int f_0(\wp)f_0(\wp,\w_1+\w_2-\wp)\, d\wp
\end{equation}
in terms of the incoming spectrum $f_0$ of each individual photon.  This entanglement is found to set a limit on the achievable fidelity of the process.

In hindsight, it is actually not difficult to envision a possible way out of this difficulty: namely, set up a situation in which momentum and energy conservation approximately apply, \emph{and} are actually not equivalent.  For this, it suffices to have the photons traveling at different velocities through the medium, or to go to a counterpropagating geometry.  These are the solutions adopted in \cite{knight} and \cite{brod1}, respectively.  Under those conditions, the two simultaneous constraints, on $k$ and $\omega$, will ultimately force $\omega \simeq \wp$ and all but eliminate the spectral entanglement in the output field (see \cite{brod2} for a more detailed discussion, and see also below).  

This still does not address point ($i$) above, which somehow seems to not apply to the system considered in \cite{jgb2010}, since this was described by a Hamiltonian model and did not require the introduction of Langevin operators.  However, close inspection of the model in \cite{jgb2010} shows that the Hamiltonian description led to a finite solution only because of the explicit introduction of a bandwidth cutoff (in agreement with the essential point made in the early works such as \cite{blow,kartner,joneckis,haus}); in other words, the field operators appearing in Eqs.~(2) and (3) of \cite{jgb2010} are not actually the free-field operators, a point made most explicitly in the subsequent work of He and Scherer \cite{BingHe}.  

A key observation, however, is that \emph{in principle} there are ways to limit a system's bandwidth without having to make it ``nonlocal in time'' in such a way as to preclude a Hamiltonian treatment.  One of these approaches, which we considered in a recent paper \cite{ViswGeaB15}, is simply to place the nonlinear medium inside a one-sided cavity.  If the medium's response time is faster than the cavity decay, then it can be treated as effectively instantaneous, and the cavity can be trusted to limit the system's effective bandwidth in a way that still allows for a fully Hamiltonian treatment \cite{jgbMOU}, as well as a strong enough interaction when the pulse bandwidth is comparable.  The approach in \cite{ViswGeaB15} was shown to work for both $\chi^{(2)}$ and $\chi^{(3)}$ processes, although, there again, spectral entanglement placed a severe limitation on the achievable fidelity.

Another bandwidth-limiting approach is to make the medium's response \emph{nonlocal in space}.  This is what was done in \cite{knight}.  As will be shown below, the spatial nonlocality extending over a distance $\sigma$ introduces a bandwidth limitation through the Fourier relation $\Delta k \sim 1/\sigma$, while still making a fully Hamiltonian description of a $\chi^{(2)}$ process possible.  (Interestingly, this approach was already shown to work with a $\chi^{(3)}$ interaction by Marzlin et al. \cite{marzlin}, albeit only under a somewhat restrictive condition.)

Finally, and for completeness, we note that point-like interactions with, for instance, atomic systems (such as those considered in \cite{brod1}), require for their description the introduction of atomic operators and do not, therefore, fall under the category of the models considered in \cite{shapiro1}, namely, pure field theories involving nothing but interacting field operators.  When the atomic systems involved start and end in their ground state, however, incoming and outgoing field states can be related by means of a unitary scattering matrix (or $S$-matrix), to which, again, the Langevin noise requirements do not apply.

\section{A $\chi^{(2)}$ medium with a spatially nonlocal response}

\subsection{General treatment} 
We consider here a $\chi^{(2)}$ (or, as in \cite{knight}, a ``$\chi^{(2)}$-equivalent'') system which in a simplified,  single-mode picture could be described by the Hamiltonian $\hat{H}=\hbar \epsilon \bigl(\hat{a}^\dagger\hat{b} \hat{c} + \hat{c}^\dagger \hat{b}^\dagger \hat{a}\bigr)$.  Starting with one $b$ and one $c$ photon, one then gets the time evolution $\ket{011} \to i\ket{100} \to -\ket{001}$.  This was proposed by Langford et al. \cite{langford} as a way to carry out a CPHASE gate and was analyzed, in a multimode treatment, by us in \cite{ViswGeaB15}, with the conclusion that spectral entanglement would prevent it from ever achieving a high fidelity.  

However, we considered only in \cite{ViswGeaB15} a copropagating arrangement with identical pulses.  Xia et al. \cite{knight}, instead, considered the case in which the $b$ and $c$ photons have different speeds, so that they pass through each other, and showed numerically that in this case high fidelities \emph{and} an overall $\pi$-phase shift were achievable.  Although the different velocities mean the $b$ and $c$ photons cannot be identical, as would be required for qubits in quantum computation, it might be possible to get around this difficulty, in principle, by using different polarizations in a birefringent medium (assuming the qubit is not encoded in the polarization state); or, perhaps more challengingly, actually shifting the frequency of the photons before they enter and after they leave the interaction medium.  

We will deal here with the most general (but still one-dimensional), multimode version of this problem, with a spatial nonlocality, as described by the following Hamiltonian:
\begin{align} \label{e3}
\hat{H} &= \hat{H}_{0} + \hat{H}_{int} \cr
\hat{H}_{0} &= \hbar v_{a} \int dk \, k\, \hat{a}^{\dagger}_{k} \hat{a}_{k} + \hbar v_{b} \int dk \, k \, \hat{b}^{\dagger}_{k} \hat{b}_{k} + \hbar v_{c} \int dk \, k \, \hat{c}^{\dagger}_{k} \hat{c}_{k} \cr
\hat{H}_{int} &= \hbar \epsilon \int_{0}^{L} dz_{a} \int_{0}^{L} dz_{b}\int_{0}^{L} dz_{c} \, f(z_{a},z_{b},z_{c}) \cr
&\quad\times \hat{A}^{\dagger}(z_{a})\hat{B}(z_{b}) \hat{C}(z_{c}) + H.c,
\end{align}
where $\hat{H}_{0}$ is the Hamiltonian of the free field and $\hat{H}_{int}$ represents the interaction with the $\chi^{(2)}$ medium. Here we consider the most general case where the $a$, $b$ and $c$ photons travel with different speeds viz. $v_{a}$, $v_{b}$ and $v_{c}$, respectively. The medium of interaction has a length $L$, but this will not figure in our calculations, since we will let the pulses pass through each other and assume the interaction starts well after they enter the medium and ends well before they leave.  The function $f(z_{a}, z_{b},z_{c})$ characterizes the nonlocal response of the medium, and following Xia et al. we will assume for it a form
\begin{equation}
f(z_{a}, z_{b},z_{c}) =   h(z_{a}-z_{b}) \ h(z_{a}-z_{c})
\label{ne4}
\end{equation}
in terms of a suitable real function $h(z)$, which we expect to be maximum at $z=0$. The operators $\hat{A}(z_{a})$, $ \hat{B}(z_{b})$ and $\hat{C}(z_{c})$ are defined as 
\begin{align} \label{e4}
 \hat{A}(z_{a}) &= \frac{1}{\sqrt{2\pi}} \int dk\ e^{ikz_{a}} \ \hat{a}_{k} \cr 
  \hat{B}(z_{b}) &=   \frac{1}{\sqrt{2\pi}} \int dk \ e^{ikz_{b}} \ \hat{b}_{k} \cr 
 \hat{C}(z_{c}) &=  \frac{1}{\sqrt{2\pi}} \int dk\  e^{ikz_{c}} \ \hat{c}_{k}
\end{align}
and  satisfy the canonical commutation relations: $[\hat{A}(z),\hat{A}^{\dagger}(z')]= [\hat{B}(z),\hat{B}^{\dagger}(z')]= [\hat{C}(z),\hat{C}^{\dagger}(z')] = \delta(z-z')$.

We shall work out this problem in the Schr\"{o}dinger picture. In the position representation, we can write the field state as
\begin{align} \label{e2} 
| \psi (t)\rangle = &\int dz_a \, \phi_{a}(z_a,t) \hat{A}^{\dagger}(z_a) \ket 0 \cr
&+ 
 \int dz_b \int dz_c \, \phi_{bc}(z_b, z_c,t) \hat{B}^{\dagger}(z_b)\hat{C}^{\dagger}(z_c) \ket 0,
\end{align}
from which we get the following equations for the ``wavefunctions'' $\phi_a$ and $\phi_{bc}$:
\begin{align}
&\frac{\partial \phi_a}{\partial t} + v_a \frac{\partial \phi_a}{\partial z_a} = -i \epsilon \int\int f(z_a,z_b,z_c)\phi_{bc}(z_b,z_c,t)\, dz_b\, dz_c \cr
&\frac{\partial \phi_{bc}}{\partial t} + v_b \frac{\partial \phi_{bc}}{\partial z_b}  + v_c \frac{\partial \phi_{bc}}{\partial z_c} = -i \epsilon \int  f(z_a,z_b,z_c)\phi_{a}(z_a,t)\, dz_a 
\label{ne6}
\end{align}
Inspection readily shows these equations to be identical to the ones considered by Xia et al.   We have, therefore, shown that their results are compatible with a Hamiltonian formalism involving only field operators satisfying the canonical commutation relations.

We will solve the system (\ref{ne6}) by working in the momentum representation, where the state of the field can be written as  
\begin{align} \label{e2} 
| \psi (t)\rangle = &\int dk_{1} \, \xi_{a}(k_{1},t) \,  \hat{a}^{\dagger}(k_{1}) \ket 0 \cr
&+\int dk_{2} \int dk_{3} \, \xi_{bc}(k_{2}, k_{3},t)  \hat{ b}^{\dagger}(k_{2})   \hat{c}^{\dagger}(k_{3}) \ket 0
\cr
\end{align}

\begin{widetext}
The Schr\"odinger equation then yields the following differential equations for the functions representing the $a$ and $b-c$ pulses, in momentum space:
\begin{align} \label{e5}
\left( \frac{\partial}{\partial t} + i k_{a} v_{a} \right) \tilde{\xi}_{a}(k_{a},t) &= - i \epsilon \sqrt{2 \pi} \int dk_{b} \ \tilde{h}(k_{b})\  \tilde{h}(k_{a}-k_{b})\  \tilde{\xi}_{bc}(k_{b},k_{a}-k_{b},t) \cr
\left( \frac{\partial}{\partial t} + i k_{b} v_{b} + i k_{c} v_{c} \right) \tilde{\xi}_{bc}(k_{b},k_{c},t) &= - i \epsilon \sqrt{2 \pi} \ \tilde{h}^{*}(k_{b}) \tilde{h}^{*}(k_{c}) \ \tilde{\xi}_a(k_{b}+k_{c},t) 
\end{align}
where Eq.~(\ref{ne4}) has been used, and $\tilde h(k)$ is the Fourier transform of $h(z)$.

We solve this system of differential equations by the method of Laplace transform. The Laplace transform of the system (\ref{e5}) is given by
\begin{subequations}
\begin{align} 
(s +  i k_{a} v_{a}) \tilde{\xi}_{a}(k_{a},s) - \xi_{a}(k_{a},0)  &= - i \epsilon \sqrt{2 \pi} \int  dk_{b}\,  \tilde{h}(k_{b})  \tilde{h}(k_{a}-k_{b}) \tilde{\xi}_{bc}(k_{b}, k_{a}-k_{b},s) \label{ne10a} \\
(s +  i k_{b} v_{b} +  i k_{c} v_{c}) \tilde{\xi}_{bc}(k_{b},k_{c},s) - \xi_{bc}(k_{b}, k_{c},0) &= - i \epsilon \sqrt{2 \pi} \tilde{h}^{*}(k_{b})  \tilde{h}^{*}(k_{c}) \tilde{\xi}_{a}(k_{b} + k_{c}, s) \label{ne10b} 
\end{align}
\label{ne10}
\end{subequations}

On substituting for $\tilde{\xi}_{a}(k_{b}+k_{c},s)$ in Eq.~(\ref{ne10b}) in terms of $\tilde{\xi}_{bc}$ using Eq.~(\ref{ne10a}) and furthermore setting $\xi_{a}(k_{a},0) = 0$, since there is no $a$ photon at $t=0$, we obtain
%
%
\begin{equation}
\tilde{\xi}_{bc}(k_{b},k_{c},s) = \frac{\xi_{bc}(k_{b},k_{c},0)}{s + ik_{b}v_{b} + ik_{c}v_{c}} - \frac{2 \pi \epsilon^{2}}{s + i(k_{b} + k_{c})v_{a}}  \frac{\tilde{h}^{*}(k_{b})\tilde{h}^{*}(k_{c})}{s + i k_{b}v_{b} + ik_{c}v_{c}} \int dk\, \tilde{h}(k) \tilde{h}(k_{b}+ k_{c}-k) \tilde{\xi}_{bc}(k,k_{b}+k_{c}-k,s)
\label{ne11}
\end{equation}

The next step is to evaluate the integral on the right hand side of Eq.~(\ref{ne11}). This can be done by shifting to dummy arguments in the same equation, viz. $k_{b} \rightarrow k' $ and $ k_{c} \rightarrow k_{b} + k_{c} - k' $ and multiplying throughout by $ \tilde{h}(k') \tilde{h}(k_{b}+k_{c}-k')$, and finally integrating both sides of the equation over $k'$. Following this procedure, we obtain
\begin{align} \label{e9}
\int dk\, \tilde{h}(k) \tilde{h}(k_{b}+ k_{c}-k) \tilde{\xi}_{bc}(k,k_{b}+k_{c}-k,s) = &\left(1+ \frac{2 \pi \epsilon^{2}}{s +i(k_{b}+k_{c})v_{a}}\int dk \frac{|\tilde{h}(k)|^{2}|\tilde{h}(k_{b}+k_{c}-k)|^{2}}{s + ikv_{b}+ i(k_{b}+k_{c}-k)v_{c}} \right)^{-1} \cr
&\times{\int dk \frac{\tilde{h}(k) \tilde{h}(k_{b}+k_{c}-k)}{s + ikv_{b}+ i(k_{b}+k_{c}-k)v_{c}}\xi_{bc}(k,k_{b}+k_{c}-k,0)}
\end{align}
On substituting Eq.~(\ref{e9}) in Eq.~(\ref{ne11}), we obtain the following expression for $\tilde{\xi}_{bc}(k_{b},k_{c},s)$.
\begin{align} 
\tilde{\xi}_{bc}(k_{b},k_{c},s) = &\frac{\xi_{bc}(k_{b},k_{c},0)}{s+ik_{b}v_{b}+ik_{c}v_{c}} - \frac{2 \pi \epsilon^{2}}{s + i(k_{b}+k_{c})v_{a}}\  \frac{\tilde{h}^{*}(k_{b}) \tilde{h}^{*}(k_{c})}{s + ik_{b}v_{b}+ik_{c}v_{c}} \cr
&\times \left(1 +  \frac{2 \pi \epsilon^{2}}{s + i(k_{b}+k_{c})v_{a}} \int dk  \frac{|\tilde{h}(k)|^{2} |\tilde{h}(k_{b}+k_{c}-k))|^{2}}{s + ikv_{b}+ i(k_{b}+k_{c}-k)v_{c}} \right)^{-1} \int dk \frac{\tilde{h}(k) \tilde{h}(k_{b}+k_{c}-k)\xi_{bc}(k, k_{b}+k_{c}-k,0)}{s + ikv_{b}+ i(k_{b}+k_{c}-k)v_{c}} \cr
\label{e10}
\end{align}
\end{widetext}
Equation (\ref{e10}) is the formal solution to our problem.  Of course, inverting the Laplace transform is generally impossible, but we are not interested in the detailed time evolution, only in the asymptotic state of the $b,c$ wavepacket after the interaction is over (formally, as $t\to\infty$).  In past work, such as \cite{GeaBNeme14,ViswGeaB15}, we have used the final value theorem of operational calculus in the form $ \lim_{t \to \infty} \xi_{bc}(k_{b},k_{c},t)=  \lim_{s \to 0} s\,  \tilde\xi_{bc}(k_{b},k_{c}, s)$, but this result is not quite applicable here.  The reason is that, in the absence of interaction, the system (\ref{e5}) does not evolve towards a constant value, but rather one has $\xi_{bc}(k_{b},k_{c},t)=  \exp(-i(k_b v_b + k_c v_c)t)\xi_{bc}(k_{b},k_{c},0)$.  In the presence of the interaction, we expect that we can separate the changing phase factor from the  slowly-varying spectral amplitude as follows:
\begin{widetext}
\begin{equation}
 \lim_{t \to \infty} \left[e^{i(k_b v_b + k_c v_c)t} \xi_{bc}(k_{b},k_{c},t)\right] =  \lim_{s \to 0} s\ \tilde\xi_{bc}(k_{b},k_{c}, s-ik_{b}v_{b}-ik_{c}v_{c})
\label{ne14}
\end{equation}
Accordingly, we make the substitution $s \rightarrow s-ik_{b}v_{b}-ik_{c}v_{c}$ in Eq.~(\ref{e10}) and take the limit (\ref{ne14}) to obtain 
%
%
%
%
%
\begin{equation} \label{e12}
 \lim_{t \to \infty} \left[e^{i(k_b v_b + k_c v_c)t} \xi_{bc}(k_{b},k_{c},t)\right] = \xi_{bc}(k_{b},k_{c},0) - 
 \frac{2 \pi \epsilon^{2} \tilde{h}^{*}(k_{b}) \tilde{h}^{*}(k_{c})}{ik_{b}(v_{a}-v_{b}) + i k_{c}(v_{a}-v_{c}) + 2 \pi \epsilon^{2} I_2(k_b,k_c)}\, I_1(k_b,k_c)
\end{equation}
\end{widetext}
where we have defined
\begin{align}
I_1 &\equiv \lim_{s \to 0} \int \ dk \frac{\tilde{h}(k) \ \tilde{h}(k_{b}+k_{c}-k)}{s + i(k-k_{b})(v_{b}-v_{c})} \, \xi_{bc}(k, k_{b}+k_{c}-k,0) \cr
I_2 &\equiv \lim_{s \to 0} \int dk \frac{|\tilde{h}(k)|^{2} |\tilde{h}( k_{b}+k_{c}-k)|^{2}}{s+i(k-k_{b})(v_{b}-v_{c})}
\label{ne16}
\end{align}
We can actually simplify the first integral in (\ref{ne16}) substantially with some straightforward assumptions. First, without loss of generality we will assume that the $b$ photon starts behind the $c$ photon and travels with a higher speed than $c$ photon. We shall denote the initial position of the center of the $b$ wavepacket by $-z_{0}$, which we take to be a large negative number. We shall also assume that the initial state is factorizable, and hence we write
\begin{equation}
\xi_{bc}(k, k_{b}+k_{c}-k,0) = e^{i k z_{0}}\xi_{b}(k,0) \ \xi_{c}(k_{b}+k_{c}-k,0)
\end{equation}
where $\xi_b(k_b,0)$ and $\xi_c(k_c,0)$ are the Fourier transforms of wavepackets centered around $z_b=0$ and $z_c=0$, respectively.  Then, making use of
\begin{equation}
\frac{1}{s+ i(k-k_{b})(v_{b}-v_{c})}=  \int_{0}^{\infty} dt \ e^{-(s + i (k-k_{b})(v_{b}-v_{c}))t}
\end{equation}
we rewrite $I_1$ as
\begin{align} \label{ne19}
I_1 = & \lim_{s \to 0}\int_{0}^{\infty} dt \ e^{-st} \ e^{i k_{b} v_{bc} t } \int_{-\infty}^{\infty}dk \, e^{ik(z_{0}- v_{bc} t)}  \cr
&\times\tilde{h}(k) \tilde{h}(k_{b}+k_{c}-k) \xi_{b}(k,0) \xi_{c}(k_{b}+k_{c}-k,0).\cr
\end{align}
where, to save space, we have introduced $v_{bc} \equiv v_b-v_c$.  Since, physically, we want $z_0$ to be much greater than the width of the wavepackets and the range of the medium's nonlocal response (width of the function $h(z)$), it is clear that the integral over $k$ in (\ref{ne19}) represents a function of $t$ that peaks around $t=t_0 \equiv z_0/v_{bc}$ and vanishes (to an arbitrarily good approximation) for both $t \gg t_0$ and $t<0\ll t_0$.  Assuming the decay for  $t \gg t_0$ is exponential or faster, we can then, first, take the limit $s=0$ (since the $e^{-st}$ factor becomes irrelevant as soon as $s$ is much smaller than both $1/t_0$ and the wavepackets' decay constant), and then formally extend the integral over $t$ to minus infinity.  Then  the integral over $t$ in (\ref{ne19}) produces a delta function, $2\pi\delta[(k_{b}-k)v_{bc}]/v_{bc}$, and we end up with
\begin{equation} \label{e14}
 I_1(k_b,k_c) =  \frac{2 \pi}{v_{bc}}\tilde{h}(k_{b})\, \tilde{h}(k_{c}) \, \xi_{bc}(k_{b},k_{c},0)
\end{equation}
%
%
%
For the second integral, a partial simplification is possible by using a well-known result involving Cauchy's principal value: 
\begin{align}  \label{ne21}
I_2 
&= \frac{\pi}{v_{bc}} \int dk \ \delta((k-k_b)v_{bc}) \ |\tilde{h}(k)|^{2} |\tilde{h}(k_{b}+k_{c}-k)|^{2} \cr
& \qquad - \frac{i}{v_{bc}} \underbrace{P \int dk \ \frac{|\tilde{h}(k)|^{2} |\tilde{h}(k_{b}+k_{c}-k)|^{2}}{k-k_{b}}}_{I_{p}} \cr
&= \frac{\pi}{v_{bc}}|\tilde{h}(k_{b})|^{2} |\tilde{h}(k_{c})|^{2} - \frac{i}{v_{bc}}I_{p},
\end{align}
where $P$ stands for the principal value and $I_{p}$ is the notation to denote this integral for brevity.  An additional advantage of (\ref{ne21}) is that it shows explicitly the real and imaginary parts of the result. 

On substituting Eqs.~(\ref{e14}) and (\ref{ne21}) in Eq.~(\ref{e12}) and carrying out some mathematical manipulation, we obtain a compact form for the final state written as
\begin{equation} \label{e17}
\xi_{bc}(k_{b},k_{c}, t \rightarrow \infty) = e^{-i(k_b v_b + k_c v_c)t} \xi_{bc}(k_{b},k_{c},0) \, e^{2i \theta(k_{b},k_{c})},
\end{equation}
where the first phase factor is just the free evolution, and the second one is the phase arising from the interaction:
\begin{equation} \label{e18}
\theta(k_{b},k_{c})= \tan^{-1}\left( \frac{2  \pi^{2} \epsilon^{2}\, |\tilde{h}(k_{b})|^{2} |\tilde{h}(k_{c})|^{2}}{\left[k_{b}v_{ab} + k_{c}v_{ac}\right]v_{bc}-2 \pi \epsilon^{2} I_{p}} \right)
\end{equation}
where we have introduced the other two relative velocities, $v_{ab} \equiv v_a-v_b$ and $v_{ac} \equiv v_a-v_c$.  In order to get a $\pi$ shift with high fidelity, we want $\theta \simeq \pm \pi/2$, to a good approximation, for all relevant values of  $k_b, k_c$.  It is, perhaps, easiest to see how that can be accomplished by considering a specific example, as in the following subsection.

\subsection{Special case: Gaussian pulses and medium response}
We shall now consider a special case where the response function of the medium is Gaussian and the initial state is also a Gaussian pulse. For this case the response function in real space is written as 
\begin{align}
f(z_{a},z_{b},z_{c}) &= h(z_{a}-z_{b})\, h(z_{a}-z_{c}) \cr
&= \frac{1}{\sqrt{\pi \sigma^{3}}}\,  e^{-(z_{a}-z_{b})^{2}/2 \sigma^{2}}\, e^{-(z_{a}-z_{b})^{2}/2 \sigma^{2}}.
\label{ne24}
\end{align} 
In  momentum space, we have 
\begin{equation}
\tilde{h}(k) = \left(\frac{\sigma}{\pi}\right)^{1/4}  \, e^{-k^{2} \sigma^{2}/2}
\label{ne25}
\end{equation}
where $\sigma$ is the length scale of medium nonlocality. The initial state is written as 
\begin{equation}
\xi_{bc}(k_{b},k_{c},0) = \frac{\sigma_{0}}{\sqrt{\pi}} \, e^{i k_{b}z_{0}}\, e^{-k_{b}^{2} \sigma_{0}^{2}/2} \, e^{-k_{c}^{2} \sigma_{0}^{2}/2}.
\label{ne26}
\end{equation}

For a Gaussian response function, the principal value integral in (\ref{ne21}) is just proportional to the Hilbert transform of a Gaussian, which can be expressed in terms of the error function of imaginary argument as
\begin{equation}
I_p = -\sigma\,e^{-\sigma^2(k_b^2+k_c^2)}\text{erfi}\left(\frac{\sigma(k_{b}-k_{c})}{\sqrt{2}}\right)
\end{equation}
%
%
%
We can then rewrite Eq.~(\ref{e18}) as
\begin{widetext}
\begin{equation} \label{e20}
\theta(k_{b},k_{c}) = \cot^{-1} \left(\frac{[k_{b}v_{ab} +  k_{c}v_{ac}]v_{bc} }{2 \pi \epsilon^{2} \sigma}\,e^{\sigma^{2}(k_{b}^{2} + k_{c}^{2})} + \text{erfi}\left (\frac{\sigma(k_{b}-k_{c})}{\sqrt{2}}\right) \right).
\end{equation}
To get a $\pi$ phase shift, we want the argument of the inverse cot function to be close to zero for all the relevant $k_b$, $k_c$.  Noting that $\xi_{bc}$ in Eq.~(\ref{ne26}) restricts $|k_b|, |k_c|$ to be not much greater than $1/\sigma_0$, we see that the argument of the erfi function goes as $\sigma/\sigma_0$, and so  $\sigma_0 \gg \sigma$ can make that term negligible, as well as make the exponential in the first term $\simeq 1$.  Then, to make the first term small enough, we require $|\Delta v|^2\ll 2\pi \epsilon^2 \sigma\sigma_0$, where $\Delta v$ is a characteristic velocity difference.  Note that we cannot simply make $v_b = v_c$, because it would nullify the whole derivation; basically, the $b$ photon would never catch up with, and interact with, the $c$ photon.  Similarly, we cannot completely eliminate the nonlocality (formally, let $\sigma \to 0$), because then there would be no way to keep the first term in (\ref{e20}) small.  

As in previous work, we  define the fidelity $F$ and the phase shift $\phi$  by the overlap between the initial and final state, i.e. $ \sqrt{F} e^{i\phi} \equiv \langle \psi(0)| \psi(t) \rangle$. In this case, to remove the phase factor of $e^{-i(k_b v_b + k_c v_c)t}$, we substitute the freely-evolving state for $\ket{\psi(0)}$ in the above expression.  In the Gaussian medium and Gaussian state considered here, we can write
\begin{align} \label{e21}
\sqrt{F}  & e^{i\phi} = \int_{-\infty}^{\infty} dk_{b} \int_{-\infty}^{\infty} dk_{c} \ \xi_{bc}^{*}(k_{b},k_{c},0) \, e^{i(k_b v_b + k_c v_c)t} \xi_{bc}(k_{b},k_{c}, t \rightarrow \infty) \cr
 &= 1- 4  \int_{-\infty}^{\infty} dk_{b}^\prime \int_{-\infty}^{\infty} dk_{c}^\prime \frac{e^{-(\tau^{2} +1)({k_{b}^\prime}^{2} + {k_{c}^\prime}^{2})}}{i \tau \alpha \left(k_{b}^\prime v_{ab}/v_{ac} + k_{c}^\prime\right)+ 2 \pi e^{-\tau^{2}({k_{b}^\prime}^{2} + {k_{c}^\prime}^{2})}\text{erfc}\left [-i \tau(k_{b}^\prime-k_{c}^\prime)/\sqrt{2}\right]},
\end{align}
where\ $\alpha \equiv v_{ac} v_{bc}/\epsilon^{2} \sigma^{2} $ is a dimensionless parameter and $\tau \equiv \sigma/ \sigma_{0}$. On the second line of Eq.~(\ref{e21}), $k_b^\prime \equiv k_b\sigma_0$ and $k_c^\prime \equiv k_c\sigma_0$ are also dimensionless variables.  This scaling simplifies the numerical calculations to follow, and also makes it clearer the limit in which we may expect $\sqrt{F} e^{i\phi} \simeq -1$; namely, when $\tau \ll 1$ and $\alpha \tau \ll 1$ (we will discuss the physical meaning of these conditions in the following section). 
%
%

In the special case considered by Xia et al. \cite{knight}, the $a$ and $b$ photon are assumed to travel with the same speed, $v_a=v_b$.  All the equations above simplify in obvious ways, and, in particular, the complex fidelity becomes
%
%
%
%
%
\begin{equation} \label{e24}
\sqrt{F}  e^{i\phi} = 1- 4 \int_{-\infty}^{\infty} dk_{b}^\prime \int_{-\infty}^{\infty} dk_{c}^\prime \ \frac{e^{-(\tau^{2} +1)({k_{b}^\prime}^{2} + {k_{c}^\prime}^{2})}}{i k_{c}^\prime \tau \alpha+ 2 \pi e^{-\tau^{2}({k_{b}^\prime}^{2} + {k_{c}^\prime}^{2})}\text{erfc}\left [-i \tau({k_{b}^\prime}-k_{c}^\prime)/\sqrt{2}\right]},
\end{equation}
where now $\alpha \equiv v_{bc}^{2}/\epsilon^{2} \sigma^{2}$.
\end{widetext}

\begin{figure}
\begin{center}
\includegraphics[width=3.5in]{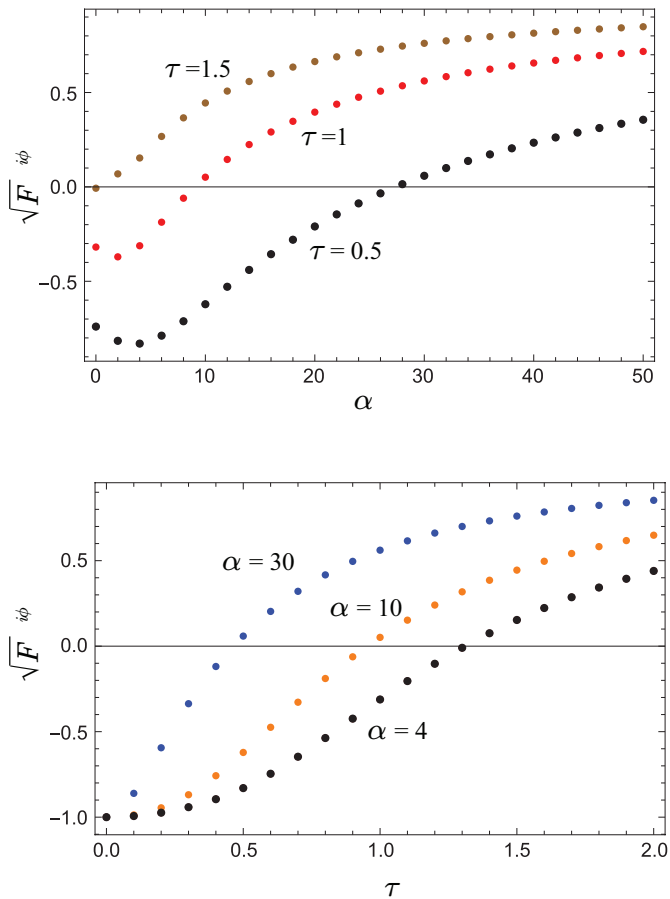}
\end{center}
\caption[example]
   { \label{fig:fig1}
Fidelity and phase shift as a function of $\alpha$ and $\tau$ for the case $v_a = v_b$.  }
\end{figure}

Figure 1 shows the result for $\sqrt{F}e^{i\phi}$ (note that $\phi$ is here limited to take on the values $0$ and $\pi$, since the quantity being evaluated is actually real) for this case, as a function of the parameters $\tau$ and $\alpha$.  This shows that it is generally more important to have a small $\tau$ than a small $\alpha$, and that, in fact, it does not matter how large $\alpha$ is, the desired result can always be achieved by making $\tau$ small enough.  Note that $\alpha$ essentially contains only medium parameters (pulse speeds, interaction strength, characteristic nonlocality length), whereas $\tau$ depends on the ``initial conditions,'' namely, the spatial extent of the pulse, $\sigma_0$.  So, what we seem to see here is that, regardless of the properties of the medium, one can always ``in principle'' get the scheme to work by making the pulse long enough.

The velocity condition $v_a = v_b \ne v_c$ of Xia et al. would be somewhat unnatural in a true $\chi^{(2)}$ medium, since in that case one would expect the $b$ and $c$ photons to be much closer in frequency to each other than they are to $a$ (in order to satisfy $\omega_a = \omega_b + \omega_c$).  The scheme of \cite{knight}, however, is in reality a four-wave mixing process with a classical pump, so one  has $\omega_p + \omega_a = \omega_b + \omega_c$, and all three $a,b,c$ photons could be very close in frequency.  Still, the result (\ref{e21}) indicates that the condition  $v_a = v_b \ne v_c$ is not really necessary.  Figure 2, computed for the set of assumptions $v_b = 1.1v_c$, $v_a = 2 v_b$, shows that the velocity of the $a$ photon does not, in fact, make any substantial difference. 

\begin{figure}
\begin{center}
\includegraphics[width=3.5in]{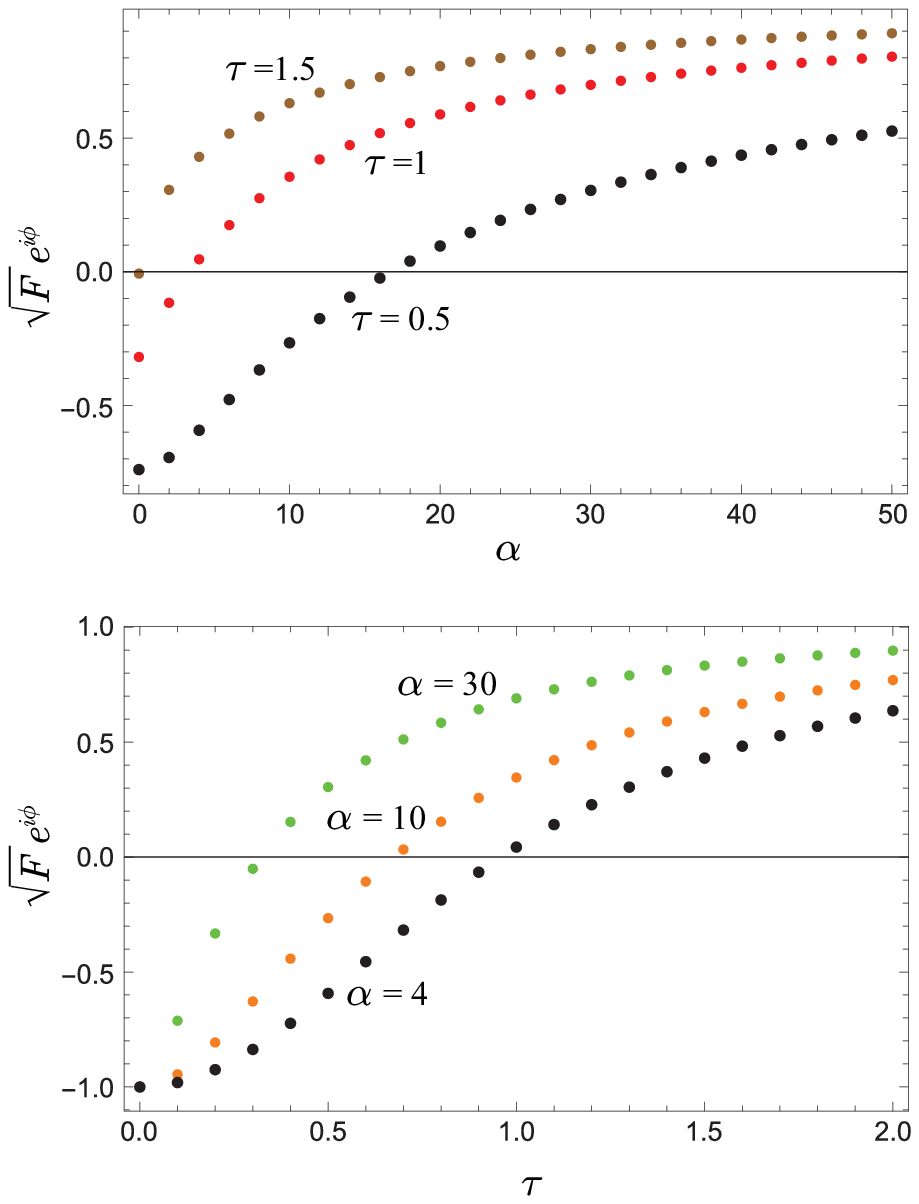}
\end{center}
\caption[example]
   { \label{fig:fig2}
Fidelity and phase shift as a function of $\alpha$ and $\tau$ for the case $v_b = 1.1v_c$, $v_a = 2 v_b$.   }
\end{figure}

\section{Discussion}


\subsection{Why increasing the pulse length helps}
As we mentioned in Section II, for copropagating pulses traveling at the same speed through a nonlinear medium increasing the pulse length actually tends to eliminate the nonlinear response altogether, because the probability that both photons  would be found within the same narrow time window (determined by the medium response time) becomes negligible.  This is clearly not the case here: since the pulses pass through each other completely, it does not matter ``where in the pulse'' each photon is initially: it is certain that they will meet eventually.

Once the photons meet, they basically have a time $t_{slip} \sim \sigma/v_{bc}$ to interact before the pulses slip past each other beyond the range of the nonlocality, $\sigma$.  Thus, the $b+c \to a$ conversion probability amplitude is proportional to $\epsilon t_{slip}$, and the corresponding probability goes as $\epsilon^2 \sigma^2/v_{bc}^2$.  However, this is only the probability assuming the photons meet in one of the roughly $N \simeq \sigma_0/\sigma$ slices into which we can (coherently) split the wavepacket.  Adding up all the probability amplitudes for these processes results in an enhancement factor of the order of $\sqrt N$ for the amplitude, or $N$ for the probability, so the overall success probability ends up being proportional to 
\begin{equation}
P \sim \frac{\epsilon^2 \sigma^2}{v_{bc}^2}\,\frac{\sigma_0}{\sigma} = \frac{1}{\alpha \tau}
\end{equation}
where $\alpha$ and $\tau$ are the dimensionless parameters introduced in the previous subsection.  This explains, qualitatively, why the scheme works in the limit $\alpha\tau \to 0$.

The factor $\sqrt N$ enhancement for a long pulse can be understood semiquantitatively as follows.  Consider the process as seen in the reference frame of the $c$ photon.  Assume for simplicity that both pulses have the same width, $\sigma_0$, and divide each of them into $N$ slices, or ``bins,'' so their state before the interaction can be written symbolically as
\begin{equation}
\frac{1}{\sqrt N} \sum_{n=1}^N \ket{z_n}_b \otimes \frac{1}{\sqrt N} \sum_{m=1}^N \ket{z_m^\prime}_c
\label{ne32}
\end{equation}
Here, $\ket{z_n}_b$ represents a state in which the $b$ photon is found in the slice centered at $z=z_n$, and similarly $\ket{z_m^\prime}_c$.  Since eventually the two photons will meet and interact for a time $t_{slip}$, all $N^2$ states in the superposition (\ref{ne32}) will, with probability amplitude $\epsilon t_{slip}$, be converted into a state that has an $a$ photon in some slice.  The width of the $a$ pulse is clearly $\sigma_0 v_{ac}/v_{bc}$, so it contains $N^\prime = N v_{ac}/v_{bc}$ slices of width $\sigma$.  Thus, on average each slice will be populated by $N^2/N^\prime$ of the terms that evolve from the superposition (\ref{ne32}), all with the same amplitude, so the $a$-pulse state can be symbolically written as
\begin{equation}
\frac{1}{N} \sum_{n=1}^{N^\prime} \frac{N^2}{N^\prime} \epsilon t_{slip}\ket{z_n}_a 
\label{ne33}
\end{equation}
The norm of this state, that is, its probability to happen, is clearly $(\epsilon t_{slip})^2 N^2/N^\prime = (\epsilon t_{slip})^2 N v_{bc}/v_{ac} = \epsilon^2 \sigma\sigma_0/v_{bc}v_{ac} = 1/\alpha\tau$, as indicated above.

Note that the conversion $b+c \to a$ is really only half the process, since what we want ultimately is to end up again with a $b,c$ pair. The eventual ``decay'' of the $a$ photon is, however, ensured as long as it stays in the nonlinear medium a sufficiently long time (which is automatically guaranteed by our formalism, since we are always taking the $t\to\infty$ limit).  More precisely, the results in the Appendix (see, in particular, Eq.~(\ref{a4})) show that the $a$ photons disappear (through the process $a\to b + c$) at a rate $\gamma = 2\pi\epsilon^2\sigma/v_{bc}$.  We therefore want $\gamma L/v_a \gg 1$, where $L$ is the length of the medium.  Since we already require $L/v_b$ (and $L/v_c$) to be greater than $\sigma_0/v_{bc}$, so the pulses have enough time to pass through each other, we see that 
\begin{equation}
\frac{\gamma L}{v_a} > \frac{2\pi\epsilon^2\sigma\sigma_0 v_b}{v_{bc}^2 v_a} = \frac{2\pi}{\alpha\tau} \frac{v_b}{v_a}\frac{v_{ac}}{v_{bc}}
\end{equation}
So as long as $v_b$ and $v_a$ are not too dissimilar, the same condition $1/\alpha\tau \gg 1$ that ensures that $b+c\to a$ happens will ensure that $a\to b + c$ happens in the medium as well.

\subsection{How the entanglement disappears}
As discussed particularly in \cite{brod2}, the spectral entanglement of the final two-photon state can be made to vanish when momentum and energy conservation give different constraints on the wavevectors of the interacting photons.  This can be seen to be the case in this system as well.

First, note that since we have assumed an effectively infinite medium, and the functions $f(z_{a}, z_{b},z_{c})$, as given by Eq.~(\ref{ne4}) exhibit translational invariance, the equations (\ref{e5}) already enforce momentum conservation.  One can see from the first Eq.~(\ref{e5}) that any momentum component $k_a$ of the $a$-photon will grow from any two $k_b$- and $k_c$-components that add up to $k_a$ (this is, in fact, the origin of spectral entanglement in these nonlinear processes).  The second Eq.~(\ref{e5}) expresses the same fact in reverse.  

The next ingredient, energy conservation, comes into play when dealing with the integrals in Eq.~(\ref{e10}), particularly the last one.  When $v_b = v_c$, this integral simply expresses the well-known spectral entanglement arising from momentum conservation, just discussed: the final momentum components at $k_b$ and $k_c$ can arise from a range of initial components $k_b^\prime$ and $k_c^\prime$ provided only $k_b^\prime + k_c^\prime = k_b+k_c$. (In Eq.~(\ref{e10}), $k_b^\prime \equiv k$ and $k_c^\prime \equiv k_b+k_c-k$.)

However, when $v_b\ne v_c$, the denominator of this integral gives us (through the pole of the Laplace transform) the long-time dependence of $\xi_{bc}$, in the form of a phase factor $\exp[i(k_b^\prime v_b + k_c^\prime v_c)t]$, and comparing it to the free-evolution factor $\exp[i(k_b v_b + k_c v_c)t]$ leads to the energy-conservation requirement $k_b^\prime v_b + k_c^\prime v_c = k_b v_b + k_c v_c$.  This can be seen in the treatment of the integral $I_1$ (Eq.~(\ref{ne16})), which is the same as Eq.~(\ref{e10}), only with the free-evolution phase factor removed: then, as discussed just below Eq.~(\ref{ne19}), the denominator of $I_1$ yields a delta function $\delta(k_b-k_b^\prime)$ in the $s\to 0$, or long-time limit.  This results in the simultaneous enforcement of energy and momentum conservation, and removes the main source of spectral entanglement in the final result (\ref{e12}).  

Alternatively, we can say that energy conservation is enforced in the long-time limit by the phase factors proportional to $v_a k_a$, $v_b k_b$, and $v_c k_c$, on the left-hand sides of Eqs.~(\ref{e5}).  This can be seen in the approximate time-domain solution to these equations presented in Appendix A.

\subsection{Role of the nonlocality}
As indicated above, we can formally get the desired high fidelity and large phase shift for any value of the nonlocality parameter $\sigma$, as long as it is not exactly zero.  It is tempting to argue that this means that the scheme is fundamentally realizable, since one would expect any real-life material or medium to exhibit \emph{some} degree of nonlocality.  Indeed, on this crucial point Xia et al. \cite{knight} cite a relatively large number of sources that mention possible nonlocal effects in four-wave mixing materials, arising from a variety of physical processes, including charge transport in a photorefractive crystal \cite{yariv}, and optical rectification in a noncentrosymmetric material that exhibits a second-order polarization in addition to the third-order one \cite{biaggio}.  

Importantly, however, we have not found in these or any of the other references cited by Xia et al. anything that can be considered a justification of the form assumed for the functions $f(z_{a}, z_{b},z_{c})$ in Eq.~(\ref{ne24}).  That is to say, the works cited allow for the possibility of nonlocal effects, in some cases (as in \cite{biaggio}) leading, indeed, to contributions to the nonlinear polarizability that depend on wavevector differences like $k_a-k_b$, but there is no indication that these effects would be \emph{the} primary bandwidth-limiting factor for the whole four-wave mixing process.  

Our tentative conclusion is that the assumed nonlocality in \cite{knight} is really just an artifice to limit the system's bandwidth in momentum space, as is needed in order to get a well-behaved field theory (as discussed in Section II above); in other words, it is not any more physically justified than the straight truncation approach we have used ourselves in other works \cite{jgb2010,ViswGeaB15}.  Clearly, in any real physical system there will be some mechanism that will limit the spread of the frequencies or wavevectors involved in the interaction; but the question that needs to be addressed for any candidate system one might propose is what exactly that mechanism is, and whether it can really be as harmless (by comparison to the temporal nonlocality discussed by Shapiro \cite{shapiro1}) as the one considered here.

\section{Conclusions}

We have carried out a detailed analytical study of the mathematical system proposed by Xia et al. that confirms their claim that a $\pi$ conditional phase shift between single photons, with arbitrarily high fidelity, is formally possible in their scheme.  Our analysis also elucidates the mechanisms that make such an outcome possible.

As a result, we must conclude that the earlier work \cite{jgb2010} by one of us was in error in considering spectral entanglement as an insurmountable obstacle to high-fidelity quantum logical operations based on single-photon nonlinear optics. The unwanted entanglement, it turns out, can be eliminated to an (in principle) arbitrary degree simply by considering a setup in which the interacting beams travel with different velocities, unlike we assumed in \cite{jgb2010,ViswGeaB15}.

Our analysis also indicates that a system like this, where the pulses pass through each other, can in principle generate large phase shifts in the very long pulse limit (in fact, it operates optimally in that limit, other things being equal).  This suggests that the noise terms associated with a potentially non-instantaneous response of the medium \cite{shapiro1} may in fact not be an insurmountable problem here, either, since their effect tends to vanish in that limit.  We note, however, that the work by Dove, Chudzicki, and Shapiro \cite{dove}, which explicitly considers this configuration in the context of a $\chi^{(3)}$ interaction, contradicts this notion, so clearly more work is required to settle this question. (It is at least conceivable that $\chi^{(2)}$ media, like the one considered here, may end up being fundamentally different from $\chi^{(3)}$ media in this regard.)

In any case, it is probably fair to say that at this point it appears increasingly likely that there may not be any fundamental limits to the realization of conditional quantum gates between single traveling photons.  Any specific proposal would need to be evaluated on its own merits, however, and the question of how it would circumvent the objections raised in previous works would have to be addressed explicitly.

\appendix

\section{Approximate time-domain solution}

In the limit that we have identified as leading to the largest fidelity ($\alpha\tau \ll 1$) it is possible to derive an approximate solution to the equations (\ref{e5}) in the time domain as follows: formally integrating the second equation and substituting in the first one, one obtains
\begin{widetext}
\begin{align}
\left( \frac{\partial}{\partial t} + i k_{a} v_{a} \right) \tilde{\xi}_{a}(k_{a},t) = &- i \epsilon \sqrt{2 \pi} \int dk_{b} \ \tilde{h}(k_{b})\  \tilde{h}(k_{a}-k_{b})\ e^{-i[k_b v_b +(k_a-k_b)v_c]t}  \tilde{\xi}_{bc}(k_{b},k_{a}-k_{b},0) \cr
&-2\pi\epsilon^2 \int_0^t dt^\prime \int dk_b |\tilde h(k_b)|^2 |\tilde h(k_a-k_b)|^2 e^{-i[k_b v_b +(k_a-k_b)v_c](t-t^\prime)} \tilde \xi_a(k_a,t^\prime)
\label{a1}
\end{align}
If we have a specific form for the functions $\tilde h$, we can evaluate the integral over $k_b$ in the second term explicitly. For the Gaussian functions we used in the body of the paper, the result is
\begin{equation}
-\epsilon^2 \sqrt{2\pi} \, e^{-k_a^2\sigma^2/2} \int_0^t dt^\prime \tilde \xi_a(k_a,t^\prime) e^{-i k_a(v_b+v_c)(t-t^\prime)/2} e^{-v_{bc}^2(t-t^\prime)^2/8\sigma^2}
\end{equation}
We can now make the approximation that $\tilde \xi_a$ is slowly-varying compared to the $\exp[-v_{bc}^2(t-t^\prime)^2/8\sigma^2]$ (which essentially only requires $\sigma$ to be small enough).  From the structure of Eq.~(\ref{a1}), it seems that a better choice for a ``slow'' function might be $e^{i k_a v_a t}\tilde \xi_a$, but this turns out not to make a difference in what follows.  Pulling $\tilde \xi_a(k_a,t)$ out of the integral, and extending the lower limit of integration to $-\infty$, this becomes
\begin{equation}
-\frac{2\pi \epsilon^2\sigma}{v_{bc}} \, e^{-k_a^2 \sigma^2(v_b^2+v_c^2)/v_{bc}^2} \,\tilde \xi_a(k_a,t) \left\{1 -i\, \text{Erfi}\left[\frac{k_a\sigma(v_b+v_c)}{\sqrt 2\, v_{bc}}\right] \right\}
\end{equation}
Now assume that $k_a \sim 1/\sigma_0$ and $\sigma \ll \sigma_0$.  We can then set the arguments of both the exponential and the error function $\simeq 0$, and we end up with the simple equation for $\tilde \xi_a$
\begin{equation}
\left( \frac{\partial}{\partial t} + i k_{a} v_{a} + \gamma \right) \tilde{\xi}_{a}(k_{a},t) \simeq - i \epsilon \sqrt{2 \pi} \int dk\ \tilde{h}(k)\  \tilde{h}(k_{a}-k)\ e^{-i(k v_{bc} + k_a v_c)t}  \tilde{\xi}_{bc}(k,k_{a}-k,0)
\label{a4}
\end{equation}
with $\gamma = 2\pi \epsilon^2\sigma/v_{bc}$. This can be integrated, with the result
\begin{equation}
\tilde \xi_a(k_a,t) = - i \epsilon \sqrt{2 \pi}\, e^{-ik_a v_c t}  \int dk\ \tilde{h}(k)\  \tilde{h}(k_{a}-k)\ \frac{e^{-i k v_{bc} t} \, \tilde{\xi}_{bc}(k,k_{a}-k,0)}{\gamma + i(k_a v_{ac} - k v_{bc})} 
\label{a5}
\end{equation}
Here we have neglected a term proportional to $e^{-(i k_a v_a + \gamma)t}$, on the grounds that this will be negligible by the time the interaction begins; again, this can be ensured for any finite $\gamma$ provided the $b$ and $c$ pulses start sufficiently far apart.  (The Fourier-transform factor $e^{-i k v_{bc} t}$ in the integral (\ref{a5}) ensures the interaction does not start, and accordingly the $a$ field does not start to grow, until pulse $b$ catches up with pulse $c$.)

One final simplification of (\ref{a5}) is possible, in the limit $\gamma \gg |k_a v_{ac} - k v_{bc}|$.  Noting that we should expect $k,k_a \sim 1/\sigma_0$, we see that this is essentially the same as the condition $\alpha\tau \ll 1$ discussed in the main text.  We conclude
\begin{equation}
\tilde \xi_a(k_a,t) \simeq - i \frac{v_{bc}}{\sqrt{2 \pi}\, \epsilon \sigma} \,  e^{-ik_a v_c t} \int dk\ \tilde{h}(k)\  \tilde{h}(k_{a}-k)\ e^{-i k v_{bc} t} \, \tilde{\xi}_{bc}(k,k_{a}-k,0) 
\label{a6}
\end{equation}
If we substitute this into the second of Eqs.~(\ref{e5}) and integrate, we conclude
\begin{align}
\tilde{\xi}_{bc}(k_b,k_c,t) \simeq &e^{-i(k_b v_b + k_c v_c)t} \Biggl[\tilde{\xi}_{bc}(k_b,k_c,0) \cr
&\quad -\frac{v_{bc}}{\sigma}\, \tilde{h}^{*}(k_{b}) \tilde{h}^{*}(k_{c}) \int dk\ \tilde{h}(k)\,  \tilde{h}(k_{b}+k_c-k)\, \int_0^t dt^\prime e^{-i (k-k_b) v_{bc} t} \, \tilde{\xi}_{bc}(k,k_b+k_c-k,0) \Biggr]
\end{align}
\end{widetext}
where again we can take the lowest limit of the time integral to $-\infty$, since the term in question is negligible before $t=0$.  
When this is done, and the long-time limit is similarly taken, one obtains a delta function $2 \pi \delta[(k-k_b)v_{bc}]$, and with the choice (\ref{ne25}) for the functions $\tilde h(k)$ (which we made use of earlier in the derivation), we get
\begin{equation}
\tilde{\xi}_{bc}(k_b,k_c,t) \simeq e^{-i(k_b v_b + k_c v_c)t} \left[1-2e^{(k_b^2+k_c^2)\sigma^2} \right]\tilde{\xi}_{bc}(k_b,k_c,0) 
\end{equation}
This again yields the desired result, $\tilde{\xi}_{bc}(k_b,k_c,t) = -e^{-i(k_b v_b + k_c v_c)t}\tilde{\xi}_{bc}(k_b,k_c,0)$, in the limit $\sigma \ll \sigma_0$.


\begin{thebibliography}{99}

\bibitem{chuang} I. L. Chuang and Y. Yamamoto, ``Simple quantum computer,'' Phys. Rev. A {\bf 52}, 3489--3496 (1995). 
\bibitem{milburn} G. J. Milburn, ``Quantum optical Fredkin gate,'' Phys. Rev. Lett. {\bf 62}, 2124 (1989).
\bibitem{shapiro1} J. H. Shapiro, ``Single-photon Kerr nonlinearities do not help quantum computation,'' Phys. Rev. A {\bf 73}, 062305 (2006).
\bibitem{shapiro2} J. H. Shapiro and M. Razavi, ``Continuous-time cross-phase modulation and quantum computation,'' New J. Phys. {\bf 9}, 16 (2007).
\bibitem{dove} J. Dove, C. Chudzicki, and J. H. Shapiro, ``Phase-noise limitations on single-photon cross-phase modulation with differing group velocities,'' Phys. Rev. A {\bf 90}, 062314 (2014).
\bibitem{jgb2010} J. Gea-Banacloche, ``Impossibility of large phase shifts via the giant Kerr effect with single-photon wave packets,'' Phys. Rev. A {\bf 81}, 043823 (2010).
\bibitem{GeaBNeme14} J. Gea-Banacloche and N. N\'emet, ``Conditional phase gate using an optomechanical resonator,'' Phys. Rev. A {\bf 89}, 052327 (2014).
\bibitem{ViswGeaB15} B. Viswanathan and J. Gea-Banacloche, ``Multimode analysis of a conditional phase gate based on second-order nonlinearity,'' Phys. Rev. A {\bf 92}, 042330 (2015).
\bibitem{HeSimon} B. He, A. MacRae, Y. Han, A. I. Lvovsky, and C. Simon, ``Transverse multimode effects on the performance of photon-photon gates,'' Phys. Rev. A {\bf 83}, 022312 (2011).
\bibitem{BingHe} B. He and A. Scherer, ``Continuous-mode effects and photon-photon phase gate performance,'' Phys. Rev. A {\bf 85}, 033814 (2012)
\bibitem{xu} S. Xu, E. Rephaeli, and S. Fan, ``Analytic properties of two-photon scattering matrix in integrated quantum systems determined by the cluster decomposition principle,'' Phys. Rev. Lett. {\bf 111}, 223602 (2013).
\bibitem{knight} K. Xia, M. Johnsson, P. L. Knight, and J. Twamley, ``Cavity-Free Scheme for Nondestructive Detection of a Single Optical Photon,'' Phys. Rev. Lett. {\bf 116}, 023601 (2016).
\bibitem{brod1} D. J. Brod and J. Combes, ``Passive CPHASE Gate via Cross-Kerr Nonlinearities,'' Phys. Rev. Lett. {\bf 117}, 080502 (2016).
\bibitem{chudzicki} C. Chudzicki, I. L. Chuang, and J. H. Shapiro, ``Deterministic and cascadable conditional phase gate for photonic qubits,'' Phys. Rev. A {\bf 87}, 042325 (2013).
\bibitem{niu} M. Y. Niu, I. L. Chuang, and J. H. Shapiro, ``Universal Quantum Computation using Coherent $\chi^{(2)}$ Interactions,'' arXiv:1704.03431v1 (2017).
\bibitem{beck} K. M. Beck, M. Hosseini, Y. Duan, and V. Vuleti\' c, ``Large conditional single-photon cross-phase modulation,'' Proc. Nat. Acad. Sc. {\bf 113}, 9740 (2016).
\bibitem{tiarks} D. Tiarks, S. Schmidt, G. Rempe, and S. D\" urr, ``Optical $\pi$ phase shift created with a single-photon pulse,'' Sci. Adv. {\bf 2}, e1600036 (2016).
\bibitem{brod2} D. J. Brod, J. Combes, and J. Gea-Banacloche, ``Two photons co- and counterpropagating through N cross-Kerr sites,'' Phys. Rev. A {\bf 94}, 023833 (2016).
\bibitem{blow} K. J. Blow, R. Loudon, and S. J. D. Phoenix, ``Exact solution for quantum self-phase modulation,'' J. Opt. Soc. Am. B {\bf 8}, 1750--1756 (1991).
\bibitem{kartner} F. X. Kartner, L. Joneckis and H. A. Haus, ``Classical and quantum dynamics of a pulse in a dispersionless non-linear fibre,'' Quantum Opt., {\bf 4}, 379 (1992).
\bibitem{joneckis} L. G. Joneckis and J. H. Shapiro, ``Quantum propagation in a Kerr medium: lossless, dispersionless fiber,'' J. Opt. Soc. Am. B {\bf 10}, 1102--1120 (1993).
\bibitem{haus} L. Boivin, F. X. K\"artner, and H. A. Haus, ``Analytical solution to the quantum field theory of self-phase modulation with a finite response time,'' Phys. Rev. Lett. {\bf 73}, 240--243 (1994).
\bibitem{ima} H. Schmidt and A. Imamo\u glu, ``Giant Kerr nonlinearities obtained by electromagnetically induced transparency,'' Opt. Lett. {\bf 21}, 1936 (1996).
\bibitem{lukin} M. D. Lukin and A. Imamo\u glu, ``Nonlinear Optics and Quantum Entanglement of Ultraslow Single Photons,'' Phys. Rev. Lett. {\bf 84}, 1419 (2000).
\bibitem{rmp} M. Fleischhauer, A. Imamo\u glu and J. P. Marangos, ``Electromagnetically induced transparency: Optics in coherent media,'' Rev. Mod. Phys. {\bf 77}, 633 (2005).
\bibitem{jgbMOU} J. Gea-Banacloche, ``Space-time descriptions of quantum fields interacting with optical cavities,'' Phys. Rev. A {\bf 87}, 023832 (2013).
\bibitem{marzlin} K.-P. Marzlin, Z.-B. Wang, S. A. Moiseev and B. C. Sanders, ``Uniform cross-phase modulation for nonclassical radiation pulses,'' J. Opt. Soc. Am. B {\bf 27}, A36 (2010).
\bibitem{langford} N. K. Langford, S. Ramelow, R. Prevedel, W. J. Munro, G. J. Milburn and A. Zeilinger, ``Efficient quantum computing using coherent photon conversion,'' Nature {\bf 478}, 360 (2011).
\bibitem{yariv} D. Engin, M. C. Cross, and A. Yariv, ``Amplitude-equation formalism for four-wave-mixing geometry with transmission gratings,'' J. Opt. Soc. Am. B {\bf 14}, 3349 (1997).
\bibitem{biaggio} I. Biaggio, ``Nonlocal Contributions to Degenerate Four-Wave Mixing in Noncentrosymmetric Materials,'' Phys. Rev. Lett. {\bf 82}, 193 (1999).




\end{thebibliography}
\end{document}